# Highly Sensitive Label-free Biomolecular Detection Using Au-WS$_2$ Nanohybrid Based SERS Substrates

Om Prakash[1#], Abhijith T[2,3#], Priya Nagpal[4], Vivekanandan Perumal[4], Supravat Karak[2], Udai B. Singh[5], Santanu Ghosh[1*]

[1]*Nanostech Laboratory, Department of Physics, Indian Institute of Technology Delhi, New Delhi 110016, India.*

[2]*Organic and Hybrid Electronic Device Laboratory, Department of Energy Science and Engineering, Indian Institute of Technology Delhi, New Delhi 110016, India.*

[3]*Department of Nanoscience and Technology, PSG Institute of Advanced Studies, Peelamedu, Coimbatore, Tamil Nadu 641004, India.*

[4]*Kusuma School of Biological Sciences, Indian Institute of Technology Delhi, New Delhi 110016, India.*

[5]*Department of Physics, Deen Dayal Upadhyay Gorakhpur University, Gorakhpur, 273009, India.*

*\* Corresponding author email: santanu1@physics.iitd.ac.in*

\# Om Prakash and Abhijith T equally contributed to this work

**Abstract**

Recent advancements in nanotechnology have led to the development of surface-enhanced Raman spectroscopy (SERS) based rapid and low-cost technologies for ultra-sensitive label-free detection and identification of molecular analytes. Herein, we utilized the synergistic plasmonic and chemical enhancement effects of Au-WS$_2$ nanohybrids to attain the high-intensity Raman signals of targeted analytes. To develop these nanohybrids, a series of monodispersed Au nanoparticles (NPs) of varying diameters from 20 to 80 nm was chemically synthesized and successively blended with liquid-phase exfoliated WS$_2$ nano-flakes of average lateral size 90 nm. They provided a maximum enhancement factor (EF) of ~$1.80 \times 10^9$ corresponding to the characteristic peaks at 1364 cm$^{-1}$ and 1512 cm$^{-1}$ for R6G analyte molecules. Theoretical studies based on the finite-difference time-domain simulations on Au-WS$_2$ nanohybrid systems revealed a huge field-intensity enhancement with an EF of more than 1000 at the plasmonic hotspots, which was induced by the strong coupling of individual plasmon oscillations of the adjacent Au NPs upon light interactions. These electromagnetic effects along with the chemical enhancement effects of WS$_2$ nanoflakes were found to be mainly responsible for such huge enhancement in Raman signals. Furthermore, these hybrids were successfully employed for achieving highly sensitive detection of the E. coli ATCC 35218 bacterial strain with a concentration of $10^4$ CFU/mL in phosphate-buffered saline media, indicating their real capabilities for practical scenarios. The findings of the present study will indeed provide vital information in the development of innovative nanomaterial-based biosensors, that will offer new possibilities for addressing critical public health concerns.

# 1. Introduction

In the past few decades, the surface-enhanced Raman scattering (SERS) based biomolecular detection technique has evolved as one of the simple and effective approaches for rapid and precise label-free detection for minuscule amounts of molecules[1–3]. The foundation of SERS technology is the plasmon resonance found on metallic nanostructures' surfaces, which concentrates incident light into nearfield evanescent waves[4–6]. The typical SERS phenomenon primarily depends on electromagnetic (EM) enhancement to increase sensitivity, which is achieved by inducing localized surface plasmon resonances at the surface of metal nanostructures[7,8]. However, in 2D materials like graphene and transition-metal dichalcogenides (TMDs), chemical enhancement (CE) in SERS is thought to be the predominant enhancement mechanism, it is based on interface dipole–dipole interactions at the substrate as well as charge transfer resonances between the substrate and analyte[9,10]. Consequently, the likelihood of Raman scattering can be increased by several orders of magnitude using different plasmonic nanostructures such as Nobel metal nanoparticles such as gold (Au), and silver (Ag), and their nanocomposites with TMDs such as MoS$_2$, WS$_2$, etc[11–14], as they have hot patches and a rough surface that promotes a localized surface plasmonic effect, which improves analyte Raman signals[15,16]. Because of the nature of nanoparticles, SERS substrates can be easily labeled for biomedical applications with biocompatible tags[17]. Furthermore, the fingerprint spectra of analytes are specific, and sample preparation for SERS measurement is straightforward and non-invasive[18]. Each year, antibiotic-resistant bacteria cause over a million fatalities worldwide and pose a serious threat to medical care[1,19,20]. According to the Centers for Disease Control and Prevention (CDC), one of the main issues facing public health in the twenty-first century is antibiotic resistance[21–23]. In recent years for the safety of food and human health care, it is essential to detect bacteria accurately and sensitively such as Escherichia coli (E. coli), Salmonella, and other waterborne and foodborne germs in water and food[24–27].

Using a variety of anisotropic nanostructures and morphologies with closely spaced metal nanoparticles or sharp edges for surface charge confinement, published research aimed to improve and increase SERS performance[28–30]. $WS_2$ is a common and stable TMD material with a high direct band gap that holds great potential for SERS applications[31,32]. A variety of materials, excluding noble-metal nanoparticles, have been employed to create SERS-active substrates such as ZnO, $TiO_2$, and graphene for their ability to provide large surface areas to attach plasmonic nanoparticles and charge transfer process[33–36]. Numerous studies have shown that the metal and metal-TMDs nanostructures as profound SERS substrates. For example, Tadesse et al. reported a detection method using SERS[2]. Gold (Au) nanorods were used to detect pathogenic bacteria as a SERS Substrate. In another study, Song et al. used Ag-$WS_2$ as a SERS substrate [37]. The method showed a good detection limit of malachite green with an enhancement factor of $7.2 \times 10^5$. In another study, Zhai et al. used the Au nano-disk array-monolayer $MoS_2$ composite formed with the help of UV lithography, electron-beam lithography, and electron beam evaporation methods, as the SERS substrate to realize the Raman detection of crystal violet (CV) molecules [38]. However, a more simplified, cost-effective approach and further optimization are possible to construct SERS substrates to detect dye molecules and pathogens [39,40].

In this proof-of-concept work, we present a label-free, straightforward technique that uses SERS to detect dye molecules R6G and E. coli bacteria strains. We use gold (Au) and tungsten di-sulphide ($WS_2$) nanohybrids as SERS substrates. The Au NPs and Au-$WS_2$ nanohybrid were synthesized using chemical routes. Firstly, the Au nanoparticle size was optimized using Rhodamine 6G (R6G) dye molecules, and then the Au-$WS_2$ nanohybrid was employed to detect very low concentrations (up to femtomolar) of dye molecules. The morphological and optical characteristics of Au-WS2 nanohybrids were studied using transmission electron microscopy (TEM) and the ultraviolet–visible near–infrared (UV–vis–NIR) spectroscopy respectively. The finite-difference time-domain (FDTD) simulation was performed to simulate the field distribution on the SERS substrate. The high enhancement was observed by the devolved SERS substrate with a high enhancement factor for the R6G dye molecules. The liquid-exfoliated $WS_2$ nanosheets decorated with Au NPs provided a huge enhancement in Raman intensities. Furthermore, the reported data shows that the SERS can be used for the identification of E. coli bacteria. In this report, the E. coli bacterial stains were detected at a lower concentration signifies that the nanohybrid SERS substrate is capable of rapid detection of dye molecules as well as bacterial strains.

## 2. Experimental Section

### 2.1. Synthesis and characterizations of Au nanoparticles and Au-WS₂ nanohybrids

Gold chloride ($HAuCl_4$), Tri-sodium citrate (TSC), Tungsten disulfide (powder, 99.9%), deionized (DI) water and Rhodamine 6G were commercially procured and used as received for all the experiments. Au nanoparticles (NPs) were synthesized using the commercially purchased gold chloride and tri-sodium citrate (TSC) as the precursor and reducing agent, respectively[41]. A series of samples different sizes of NPs were prepared by changing the amount of TSC from 50 to 200 µl. In this synthesis, a particular amount of 38.8 mM TSC aqueous solution was quickly injected into 5 ml of boiling 0.5 mM aqueous gold chloride under stirring conditions. The colourless solution slowly changed to red/red-violet colour after 15 minutes of stirring, which indicated the formation of Au NPs. Ultra-thin two-dimensional structures of $WS_2$ were prepared using a liquid phase exfoliation method [42,43]. In this method, initially, the bulk $WS_2$ powder of 50 mg was dissolved in 10 ml of de-ionized water using a magnetic stirrer. Afterward, this solution was sonicated for 8-10 hours using a bath sonicator at a temperature below 10°C. Finally, this dispersion was centrifuged a few times at 5000 rpm to completely remove the aggregates and unexfoliated bulk materials. The yellow-coloured supernatant in the final dispersion indicated the presence of mono/few layers of $WS_2$ nanoflakes. The morphological characteristics of Au nanoparticles and Au-$WS_2$ nanohybrids were studied using TEM (JEOL JEM-1400). The UV–vis–NIR spectroscopy (Shimadzu, UV-2450) was performed to study the shift in the LSPR response with respect to the variations in the Au nanoparticle size and to study intrinsic optical properties of nanohybrids. The FDTD simulations (Ansys Lumerical) were performed to monitor the field intensity enhancement around the NPs.

### 2.2. SERS detection

The SERS measurements were carried out using the Raman spectroscopy (Renishaw plc, Micro Raman Spectrometer) using a 30 mW laser beam with an excitation wavelength of 785 nm to probe the R6G molecule. A microscope objective of 50× magnification with a 0.50 numerical aperture was used having a spectral resolution of 1 cm$^{-1}$, and the spot diameter and depth of focus were around 0.2 and 5 µm, respectively. The spectra were collected over a 1 × 1 µm² area with accumulations = 3, and exposure time = 20 s.

## 2.3. Preparation of Bacterial samples

Bacteria were revived from glycerol stocks by transferring them into 1mL of freshly prepared Luria-Bertani (LB) broth. These primary cultures were incubated overnight at 37°C with shaking at 220 rpm. Subsequently, the primary cultures were diluted in the ratio of 1:100 and allowed to grow until they reached the mid-log phase. The optical density of secondary cultures was measured at a wavelength of 600 nm and subsequently adjusted to 0.09 which corresponds to approximately $1.0 \times 10^8$ CFU/mL. These adjusted secondary cultures were serially tenfold diluted in PBS (Phosphate buffered saline) to prepare bacterial samples with concentrations of $10^4$ CFU/mL.

## 3. Results and Discussion
## 3.1. Au nanoparticle Size variation

To study the morphological and size variation of synthesized Au NPs, TEM measurements were performed, and the corresponding TEM images of Au NPs are shown in Figures 1 (a)-(d), and their size distribution histograms are shown in the inset of respective figures. The size of NPs was found to be continuously increased from 20 to 80 nm when the TSC amount reduced from 200 to 50 μl. To study the optical responses of these NPs, the UV–Vis-NIR spectra were recorded (figure 1 (e)). A systematic red-shift from 521 to 546 nm was observed in the surface plasmon resonance (SPR) peak of NPs as the TSC amount reduced from 200 to 50 μl, also the SPR broadening was observed as the size of Au NPs increased. The estimated size and SPR peak of NPs were also plotted against the TSC amount, as shown in Figure 1 (f). A sudden variation in size and SPR peak was noted as the TSC amount reduced from 100 to 50 μl, indicating that the TSC amount is critical in determining the size of Au NPs.

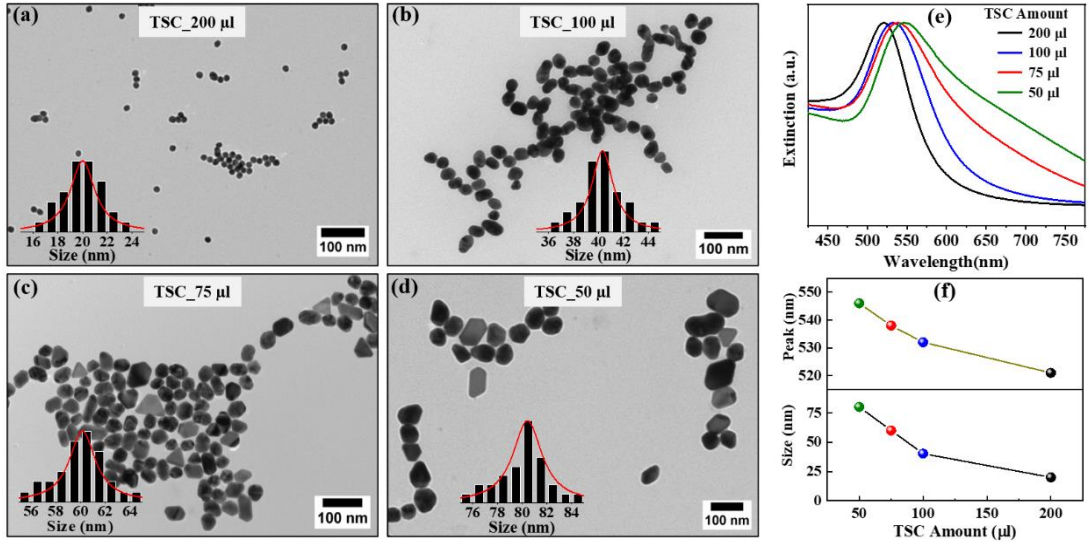

**Figure 1.** TEM images of synthesized Au NPs with TSC amount of (a) 200 µl, (b) 100 µl, (c) 75 µl, and (d) 50 µl, (e) the extinction spectra of Au NPs synthesized with different amounts of TSC, and (f) the plots showing the variations of size and SPR peak position of NPs with respect to the TSC amount.

### 3.2. Near-field intensity measurements using FDTD simulations

To estimate the near-field intensity enhancement around the Au NPs, the FDTD simulations were performed using a commercially available Ansys Lumerical FDTD solutions software. Most of the synthesized NPs appeared in spherical/quasi-spherical shapes. Therefore, all the simulations were carried out using spherical NPs. To consider the utmost realistic case, Au NPs were modelled on top of a Si substrate, as shown in Figure 2 (a). An x-polarized total-field scattered-field (TFSF) light source with a wavelength of 785 nm propagating in the (-) y direction was chosen to illuminate this model. An x-y frequency domain field-monitor was used to monitor the field-intensity enhancement profiles of Au NPs. Figure 2 (b) shows the field-intensity enhancement profiles of different sizes of Au NPs on Si substrate. A significant increase in field intensity was observed around the Au NPs when their size increases from 20 to 80 nm. To estimate the field intensity enhancement factor, $|E|^2/|E_0|^2$ (EF), the line profiles were captured across the centre region of NPs as shown in Figure 2 (c). The location of line profiles is represented by white dashed lines in the field intensity enhancement profiles shown in Figure 2 (b). All the NPs exhibited a high electric-field intensity at their surface, and this field sharply decreased to their lateral sides. The NP of size 20 nm exhibited an EF around 2 at its surface, and this EF dropped to unity at 2.5 nm away from the surface. On the other hand, the bigger NPs exhibited high EF values near their surface, and they retained significantly at several nanometres away from the surface. The NP of size 80 nm showed an EF of around 6, and more than 50% of this EF was retained over a lateral distance of 7.5 nm. These results suggest that the bigger NPs can provide a high field intensity over a widespread region, which is highly beneficial for carrying a large number of bio-analytes for surface-enhanced Raman-scattering (SERS) applications.

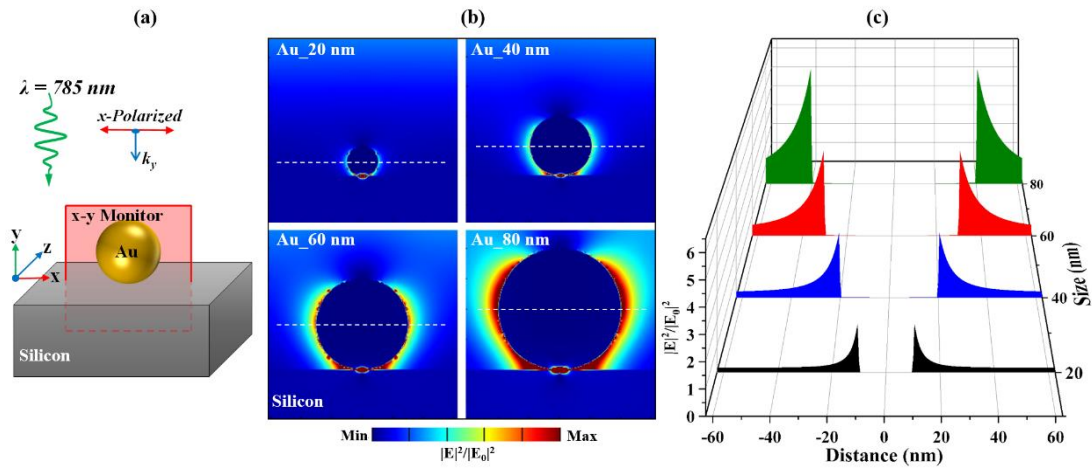

**Figure 2.** (a) FDTD simulation structure, (b) field-intensity enhancement profiles around different sizes of Au NPs, (c) the field-intensity line profiles simulated across the centres of NPs.

### 3.3. Au nanoparticle Size impact on SERS Spectra

The Au nanoparticle size change affects the SERS signal intensity as the change in Au NPs size leads to a change in the optical properties of the substrate. Enhancement factor values also differ for different-sized Au NPs. To optimize the size of Au nanoparticles for a better SERS signal, the Raman spectra were recorded using 1 μM R6G solution. For this purpose, a mixed solution of R6G dye molecules and different-sized nanoparticles was drop-casted on the Si substrate and then dried in hot air. Raman intensities of different-sized nanoparticles with R6G molecules are shown in Figure 3. The Raman spectra showed huge variations with the increase in nanoparticle size. The SERS enhancement is less in smaller Au NPs of sizes 20 nm and 40 nm. The enhancement factor was calculated as $3.20 \times 10^7$ and $9.10 \times 10^7$ for 20 nm and 40 nm respectively. The maximum enhancement is observed in the case of 60 nm Au nanoparticle. The enhancement factor was calculated as $7.40 \times 10^8$ for 60 nm. As the nanoparticle size was further increased to 80 nm, the enhancement decreased, and the enhancement factor was calculated as $6.40 \times 10^7$. All the enhancement factor was calculated corresponding to a 1512 cm$^{-1}$ characteristic peak for R6G concentration of 1 μM. In the Au nanoparticle synthesis process, the HAuCl$_4$ concentration was kept constant and TSC solution concentration was varied, due to which as the size of the nanoparticles increases the concentration of the number of the nanoparticle in the solution decreases. This may result in a decrease in the enhancement factor due to more separations between the adjacent nanoparticles (size > 60 nm) as observed in the enhancement factor calculations. In addition, Recent reports indicated that for bigger-sized Au NPs, the influence of scattering becomes considerably stronger than the absorption[44]. As a result, 60 nm Au nanoparticles were further utilized in the study for the Au-WS2 nanohybrids.

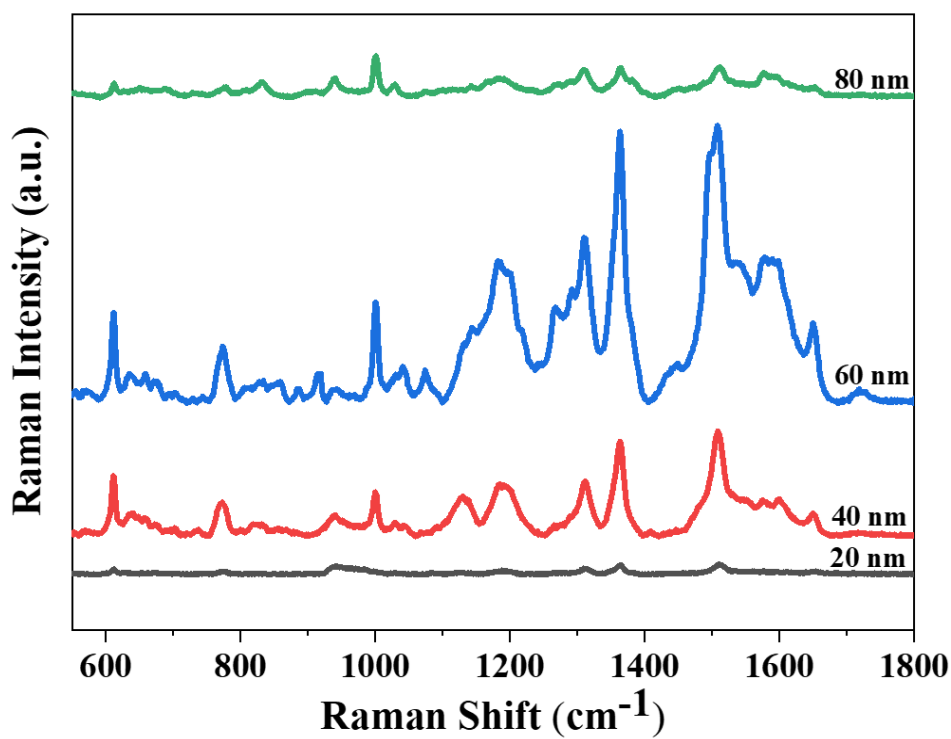

**Figure 3.** Raman spectra of R6G with different sizes of Au nanoparticles.

### 3.4. Au-WS$_2$ nanohybrids

In addition to plasmonic metal nanostructures, the researchers have explored transition-metal dichalcogenide materials to attain the Raman signal enhancement of analyte molecules, where the enhancement has been reported due to the chemical enhancement mechanism [45]. To couple the advantages of both plasmonic and chemical enhancement effects for the development of highly sensitive SERS substrates, The exfoliated WS$_2$ material was blended with synthesized Au NPs of size 60 nm by stirring their solutions for 24 hours. TEM image and size distribution histogram of exfoliated WS$_2$ is shown in Figure 4 (a), which reveals the formation of nanoflakes with lateral size of 60-120 nm. The SAED pattern shown in Figure 4 (a) also confirmed the polycrystalline nature of nanoflakes. Detailed morphological and spectroscopic characterizations of 2D WS$_2$ structures were included in our earlier reports [46–48]. As shown in the TEM image of nanohybrids in Figure 4 (b), the Au NPs appeared uniformly in the vicinity of WS$_2$ nanoflakes, which confirmed the proper blending of both nanostructures. The significant overlapping of characteristic A and B excitonic peaks of WS$_2$ nanoflakes with a plasmonic band of Au NPs further confirmed the uniform dispersion of these nanostructures in the present nanohybrid solution (Figure S1).

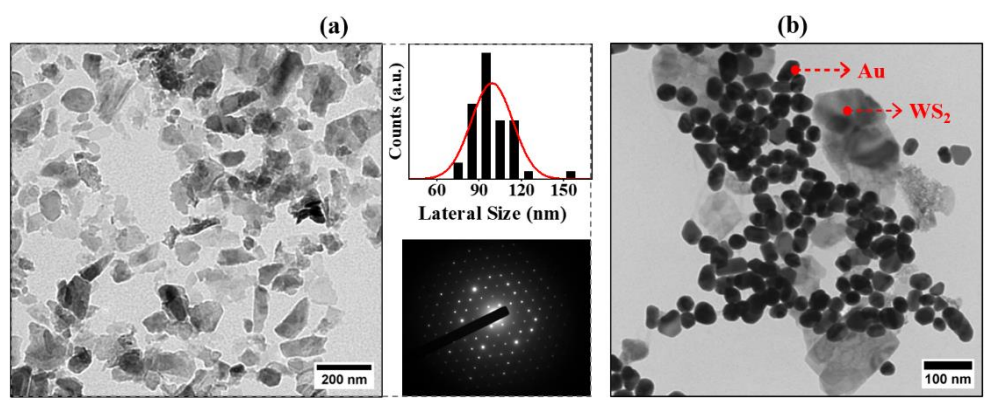

**Figure 4.** (a) TEM image, lateral size distribution histogram, and SAED patterns of WS$_2$ nanoflakes, and (b) TEM image of nanohybrids.

## 3.5. SERS of Au-WS$_2$ nanohybrids

For probing the SERS sensitivity on the Au-WS$_2$ nanohybrid substrates, R6G molecules Raman spectra of different concentrations as low as $1 \times 10^{-15}$ M were collected employing the R6G resonance excitation wavelength of 785 nm. The SERS measurements were taken corresponding to WS$_2$ Nanosheets, Au nanoparticles, and Au-WS$_2$ nanohybrids. As shown in Figure 5 (a) corresponding to the WS$_2$ nanosheets we observed a weak Raman signal as chemical enhancement contribution is less, on Au NPs the enhancement is significantly high because of electromagnetic enhancement. We observe a huge Raman signal with Au-WS$_2$ nanohybrids, this is a combined result of chemical enhancement (due to WS$_2$) and electromagnetic enhancement due to (Au nanoparticles). Furthermore, the SERS activity was checked using R6G as a probe molecule at different concentrations ranging from micromolar to femtomolar as shown in Figure 5 (b). Nanohybrid formation with WS$_2$ gives chemical enhancement contribution in addition to electromagnetic enhancement by Au NPs which leads to additional characteristics peaks of R6G at 693 cm$^{-1}$, 1310 cm$^{-1}$, and 1416 cm$^{-1}$ and a huge enhancement in overall SERS signal[49]. These results confirm that synthesized SERS substrate can be utilized for minute amounts of molecule detection.

Furthermore, Raman mapping was performed over the optimized Au-WS$_2$ SERS substrate to strengthen the evenness of the SERS signal over an area of 5 × 5 μm$^2$ in a random location with one accumulation. The exposure time was reduced to 10 sec to reduce the total time for signal collection with laser power set to 30 mW. Raman mapping corresponding to two diverse characteristic peaks (1364 and 1512 cm$^{-1}$) of the R6G molecule is presented in Figure 6 (a-b), indicating the reliability of nanohybrid substrates for SERS measurements. A high SERS signal was achieved due to the existence of hotspots which results in electric field confinement. The plot associated with Raman spectra obtained at each of the 121 pixels is displayed in Figure 6 (c), representing adequate uniformity of the substrate. The Raman peak at 1364 cm$^{-1}$ and 1512 cm$^{-1}$ of R6G was considered for character reproducibility of the SERS substrate. The standard deviation of Raman intensities for two diverse characteristic peaks 1364 and 1512 cm$^{-1}$ was calculated to be 10% and 12.5% respectively. With these mapping results, it can be undoubtedly considered that the SERS substrate is substantially uniform.

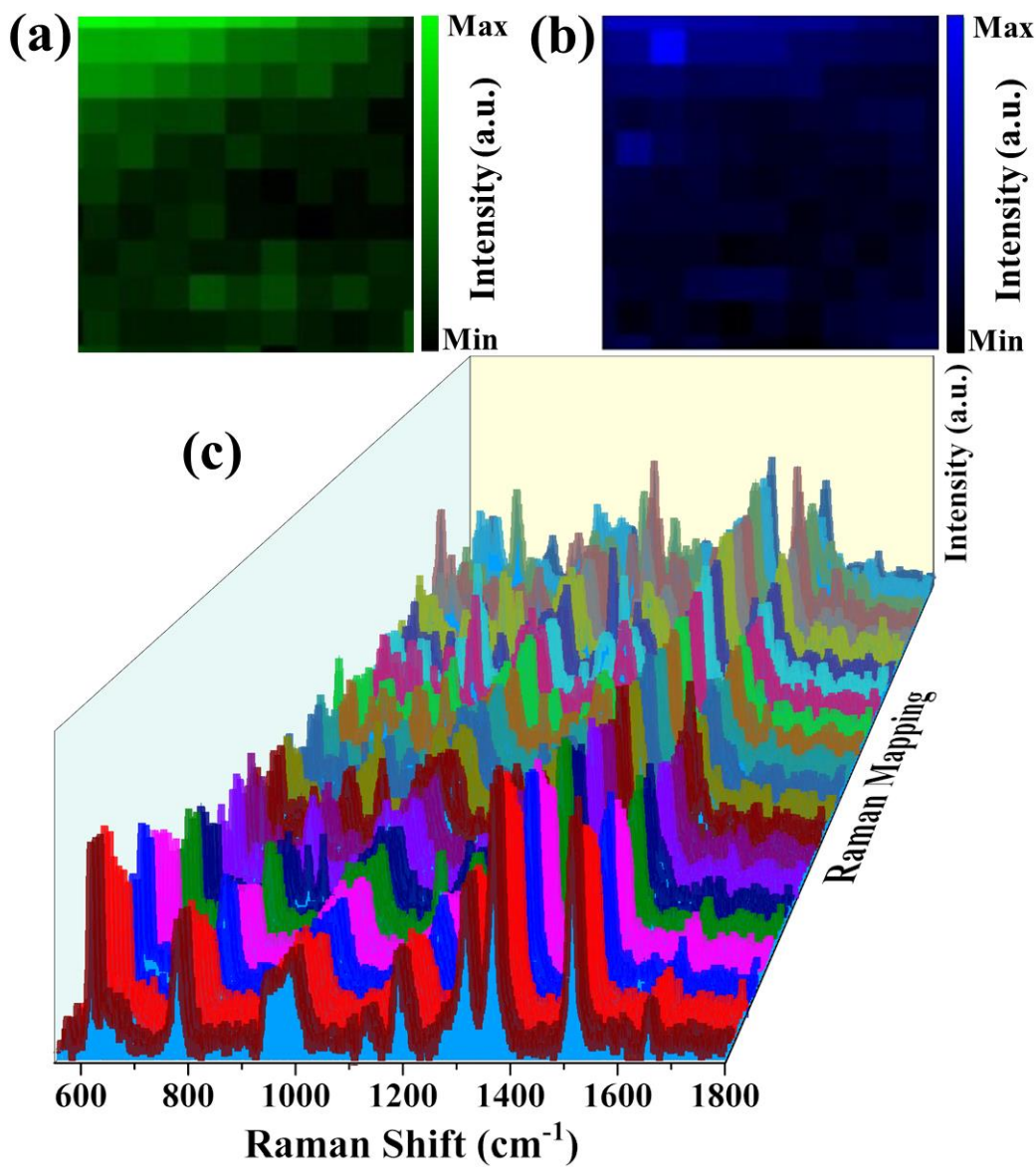

**Figure 6.** 2-D Raman mapping images of the $10^{-9}$ M R6G molecules over the substrate for (a)1364 cm$^{-1}$ and (b) 1512 cm$^{-1}$ Raman modes, respectively. (c) Raman spectra of each point in the measurement region.

### 3.6. Near-field intensity measurements at plasmonic hotspots using FDTD simulations.

In the case of nanohybrids, as Au NPs sit near each other, the surface plasmon oscillations of adjacent particles couple together, leading to the formation of plasmonic hotspots with highly localized fields between their narrow gaps, as discussed elsewhere [50]. The interaction of these hotspots with analyte molecules increases their Raman signals many times higher than that with the isolated NPs. To monitor the field intensity enhancement at the hotspot, the FDTD simulations were carried out on a simplified two-particle system with each NP size of 60 nm and an interparticle gap of 2 nm. For this purpose, a model with structure, Si/WS$_2$/Au NPs, was constructed as shown in Figure 7 (a) with similar simulation conditions as discussed in the case of Si/Au NP structure. The field intensity enhancement profile around the adjacent Au NPs is shown in Figure 7 (b). A substantial field intensity enhancement was observed in the vicinity of these NPs, mainly at the interparticle gap. To estimate the field intensity EF across horizontal and vertical directions of the hotspot, the field intensity line profiles were simulated along the white dashed lines shown in Figure 7 (b). The line profiles shown in Figure 7 (c) revealed the strong confinement of the electric field at the hotspot with an EF of more than 1000. For a particular NP system, the EF value at the hotspot crucially depends on many factors, such as the number, interparticle gap, and alignment of NPs. We strongly believe that the enhancement in Raman signals of targeted analytes was mainly due to the strong interaction of this induced plasmonic hotspot along with the chemical enhancement effects of WS$_2$ nanoflakes.

## 3.7. Enhancement Factor Calculation

The enhancement factor (EF) was calculated for the results obtained from the 1nM R6G solution drop-casted on Au-WS$_2$ and bare silicon SERS substrates, as shown in the Supporting information and Figure S2. The enhancement factor was determined using the following formula[51].

$$EF = \frac{I_{SERS} \times N_{RAMAN}}{I_{RAMAN} \times N_{SERS}} \quad (1)$$

where $N_{Raman}$ and $N_{SERS}$ are the numbers of the probe molecules absorbed on the bare silicon surface and in the Au-WS2 nanohybrid (drop casted on a silicon substrate); and $I_{Raman}$ and $I_{SERS}$ are the intensities of the vibrational mode of R6G absorbed on bare silicon and SERS substrate, respectively.

$$N_{Raman} = \frac{Ah\rho N_A}{M} \quad (2)$$

$$N_{SERS} = \frac{cVAN_A}{s} \quad (3)$$

where A is the area of the incident laser spot on the substrate surface, h is the depth of laser penetration, $\rho$ is the density of the probe molecule, and M is the molecular weight. $N_A$ is the Avogadro number. V, c and s are the volume, molar concentration, and surface area of the R6G solution, respectively. For every measurement, the experimental parameters remained constant.

The calculated EF for different Raman peaks of probe molecules with their modes of vibrations are listed in Table 1.

**Table 1.** Assigned Raman modes of R6G molecule and the enhancement factor [52]

| Sr. No. | Raman Peak (cm$^{-1}$) | Modes of vibration | Enhancement factors (EFs) |
|---|---|---|---|
| 1 | 611 | Aromatic C-C-C ring in-plane vibration | $9.25 \times 10^8$ |
| 2 | 775 | Out-of-plane C-H vibration | $8.52 \times 10^8$ |
| 3 | 1076 | In-plane C-H vibration | $7.86 \times 10^8$ |
| 4 | 1184 | In-plane bending of C-H | $9.34 \times 10^8$ |
| 5 | 1364 | C-C Aromatic stretching | $1.72 \times 10^9$ |
| 6 | 1512 | C-C Aromatic stretching | $1.80 \times 10^9$ |
| 7 | 1575 | C-C Aromatic stretching | $8.81 \times 10^8$ |

### 3.8. Bacterial pathogen detection of E. coli using SERS

Post-optimization, we acquired SERS spectra of bacteria strains using Au-WS$_2$ SERS substrate and studied the vibrational energy levels containing molecular fingerprint information. Figure 8 shows the SERS spectral characteristics of E. coli ATCC 35218, when the Raman spectra of the bacteria sample were taken on the bare Si substrate no characteristic peak was observed as shown in the figure below, whereas the Raman spectra on Au-WS$_2$ SERS substrate in which we can see the characteristic peaks corresponding to E. coli bacterial strain. The SERS spectra of E. coli ATCC 35218 exhibit characteristic peaks around 736 cm$^{-1}$ (Glycosidic ring/adenine/CH2 rocking), 1149 cm$^{-1,}$ and 1210 cm$^{-1}$ (C–C skeletal modes in proteins) 1306 cm$^{-1}$ (Wagging of CH2 of agar solution), 1438 cm$^{-1}$ and at 1494 cm$^{-1}$ (Ring stretching of guanine, adenine, –C–O vibration modes of peptidoglycan)[53–55]. These results signify that the developed SERS substrate is versatile in identifying the different molecules.

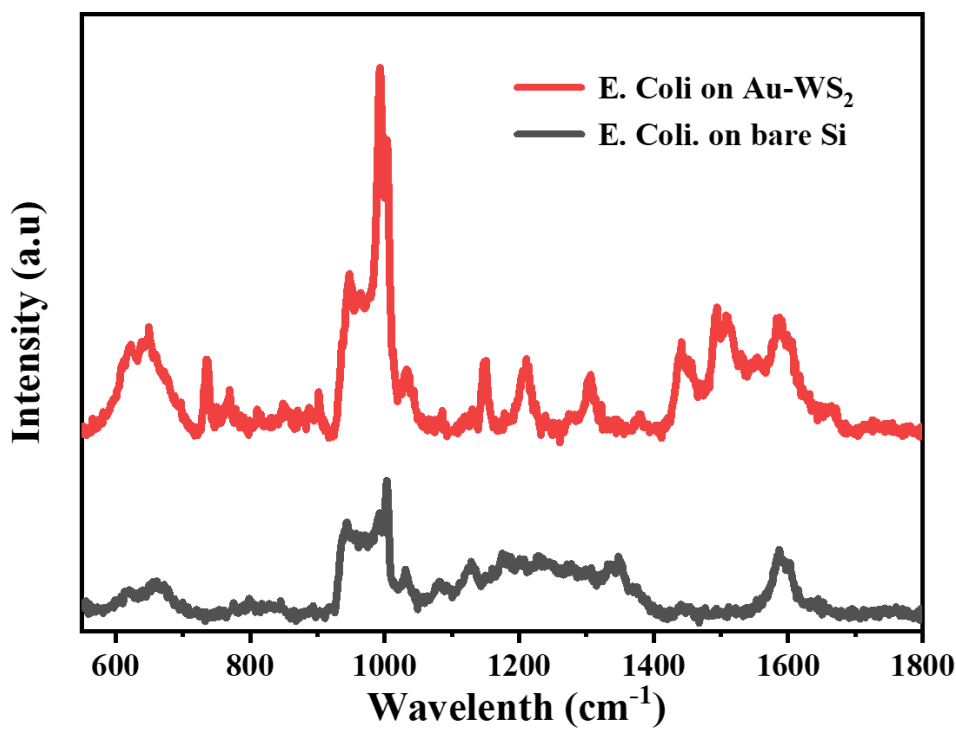

**Figure 8.** SERS spectra of Escherichia coli bacteria strain in PBS media

## 4. Conclusion

In summary, we have reported the design of a heterostructure material by employing transition metal dichalcogenide, $WS_2$, and plasmonic, Au NPs for label-free SERS detection of dye molecules and bacteria. It was found that the Au-$WS_2$ nanohybrid substrates can provide highly sensitive detection of R6G dye molecules from micromolar to femtomolar ($10^{-6}$ M to $10^{-15}$ M) concentrations. The surface plasmon resonance (SPR) peak of NPs showed a consistent red-shift from 521 nm to 546 nm as the Au NP size increased from 20 nm to 80 nm. Au NPs with a size of around 60 nm were found to be the best SERS reporters in the optimization process. Furthermore, its nanohybrid formation with $WS_2$ gives chemical enhancement contribution in addition to electromagnetic enhancement by Au NPs which leads to getting additional characteristics peaks of R6G at 693 $cm^{-1}$, 1310 $cm^{-1}$, and 1416 $cm^{-1}$ and a huge enhancement in overall SERS signal. The enhancement factor corresponding to the characteristic peak 1512 $cm^{-1}$ was found to be around $1.80 \times 10^9$. A notable increase in field intensity and the occurrence of hot spots was verified using FDTD simulated profiles. In addition, the present nanohybrid SERS substrates were able to detect E. coli bacteria at $10^4$ CFU $mL^{-1}$ in PBS media which shows their ability to be used as rapid pathogen detection substrates. This Au-$WS_2$ nanohybrid platform will exhibit significant promise as an affordable, reliable, and portable sensing platform for upcoming applications.

**Supporting Information**

**Extinction spectra of Au NPs, WS$_2$ nanoflakes, and Au-WS$_2$ nanohybrids; Enhancement factor calculation.**

**Notes**

The authors declare no competing financial interest.

**Author Information**

**Corresponding Author**

**Santanu Ghosh** - *Nanostech* Lab*, Department of Physics, Indian Institute of Technology Delhi, New Delhi 110016, India*
Email: santanu1@physics.iitd.ac.in

**Authors**

**Om Prakash-** *Nanostech Laboratory, Department of Physics, Indian Institute of Technology Delhi, New Delhi 110016, India.*

**Abhijith T-** *Organic and Hybrid Electronic Device Laboratory, Department of Energy Science and Engineering, Indian Institute of Technology Delhi, New Delhi 110016, India. & Department of Nanoscience and Technology PSG Institute of Advanced Studies Peelamedu, Coimbatore Tamil Nadu 641004, India.*

**Priya Nagpal**- *Kusuma School of Biological Sciences, Indian Institute of Technology Delhi, New Delhi 110016, India.*

**Vivekanandan Perumal** -*Kusuma School of Biological Sciences, Indian Institute of Technology Delhi, New Delhi 110016, India.*

**Supravat Karak**- *Organic and Hybrid Electronic Device Laboratory, Department of Energy Science and Engineering, Indian Institute of Technology Delhi, New Delhi 110016, India.*

**Udai B. Singh**- *Department of Physics, Deen Dayal Upadhyay Gorakhpur University, Gorakhpur, 273009, India.*

**Acknowledgments**

The author, Mr. Om Prakash, would like to thank UGC for funding the fellowship. The author would also like to thank Dr. Rajni Gaind for generously providing Cephalosporin-resistant *E. coli* ATCC 35218 (a beta-lactamase producing lab strain) bacterial strain, Department of Microbiology, Vardhaman Mahavir Medical College, and Safdarjung Hospital. The author would also like to thank the Central Research Facility, and Nanoscale Research Facility at IIT Delhi for their characterization facilities.